# OPTIMAL INTEGRATION OF HEAT-PUMP AND SOLAR THERMAL ENERGY IN THE PRE-HEATING LOOP OF WOOD AND GAS BOILER BASED DISTRICT HEATING SYSTEM

METTALI Hamza[1,2]*, ROUSSET François[1], BIDEAUX Eric[2], CLAUSSE Marc[1]

[1]INSA Lyon, CNRS, CETHIL, UMR5008, F-69621 Villeurbanne, France

[2]INSA Lyon, CNRS, ECL, AMPERE, UMR5005, F-69621 Villeurbanne, France

*Corresponding Author: hamza.mettali@insa-lyon.fr

## ABSTRACT

The integration of renewable sources is essential for decarbonizing heat production in district energy networks. Beyond biomass-based solutions, solar thermal energy, with or without heat pumps, presents a significant opportunity. However, system performance is highly dependent on outdoor and setpoint temperatures. This study aims to optimize system design using a multi-criteria approach that considers techno-economic and environmental (CO2) factors. A Mixed-Integer Linear Programming (MILP) model is developed, incorporating temperature discretization for problem linearization and capturing key dynamic characteristics of heat generators. The model improves convergence, reducing a 19% MIP gap in 26 hours to 10% in 12 hours by dissipating 6% excess solar heat. A multi-scenario analysis under two carbon taxation levels and different CO2 emission cases revealed solar integration up to 11,932 m² but increased gas reliance (50%) and TES losses (49%). Wood boiler inclusion reduced solar dependency, covering 45% of heat, lowered LCOH, but limited renewable penetration. Higher carbon taxes boosted solar adoption but faced storage inefficiencies, while biomass enhanced cost efficiency and system stability.

## 1 INTRODUCTION

Shifting to 4$^{th}$ and 5$^{th}$ generation District Heating systems (4GDHS and 5GDHS), which operates at lower temperatures, is essential for allowing more effective integration of renewable energy, especially when combined with advanced design and operation's optimization strategies (Lund et al., 2018). The optimization of these systems implies to solve problems which are of MINLP type by nature, so that a trade-off has to be found between model precision and complexity vs. reasonable computation time. (Sporleder et al., 2022). The non-linear behaviors is due to relations between different design and operational variables including mass and heat flows, temperatures (Kotzur et al., 2021) and different dynamics behavior constraints such as part-load efficiency, ramp-up/ramp-down and uptime/downtime (Wirtz et al., 2021a). In addition, for technologies such as Heat-Pumps (HP) and Solar thermal systems, non-linearty is also due to dependence of performance on operating temperatures. Some authors keep the MINLP formulation to solve the problem which requires applying a reduction on the size of the problem by relying on steady-state equations (temperatures, mass flow, …) and by neglecting dynamic load variations, e.g. (Mertz et al., 2016). Others limit the modeling to integer bilinear constraints (Hering et al., 2021). Hering et al. (2021) utilize a Mixed-Integer Quadratically Constrained Program (MIQCP) to optimize a low-temperature district heating network, incorporating heat pump placement and operational strategies while accounting for temperature-dependent efficiency. The inclusion of bilinear constraints increases computational complexity, and the fixed temperature assumption limits flexibility. Delubac et al. (2021) use NLP formulations including continuous mass flow and temperature variables, highlighting the importance to incorporate the performance of heat pumps and solar thermal in optimizing the design of DHS, especially for the efficiency of the solar thermal system, in regard to the operating temperatures. To lower the complexity of the model, MILP formulations presents an interesting alternative in mathematical optimization. Krützfeldt et al. (2021) proposes a Mixed-Integer

Linear Programming (MILP) model for designing and operating heat pump systems in residential buildings. The study demonstrates that introducing a COP depending on the supply temperature results in reduction by 20% in optimal Heat pump capacity and increase by 300% in TES volume compared to a priori calculated COP but the Taylor-approximation results on limitations on the model convergence (MIPGAP ≈50%) and accuracy (from 1.6% to 6.9% underestimation of electricity consumption to non-linear recalculations. Terlouw et al. (2023), Schweiger et al. (2017) and Dorotic et al. (2019) expanded this by proposing a MILP-based multi-objective optimization frameworks with includes integrating renewable heating sources in DHS considering the performance sensitivity to a constant operating temperature (Dorotić et al., 2019; Schweiger et al., 2017; Terlouw et al., 2023a, 2023b). Another MILP-based approach is presented by Wirtz et al. (2021), who optimize 5th-generation district heating and cooling (5GDHC) networks. The model incorporates temperature discretization to account for nonlinear relationships between network temperature and heat pump performance. Implemented within a Model Predictive Control (MPC) framework, the optimization improves cost efficiency by dynamically adjusting operational strategies over a foresight horizon. (Wirtz et al., 2021b).

From this literature review, it appears that the inclusion of temperature as a decision variable in a MILP formulation for a design problem has been scarcely studied so far. Hence, this paper aims to optimize the design of a district heating combining several renewables heat sources, the performance of two of them, solar thermal loop & heat-pump being temperature dependent, using a MILP formulation with temperature discretization. To achieve this, a relaxation function is proposed to facilitate the convergence for outlet panel temperature of the solar loop which is challenging for the optimization problem because the solar resource intermittency. Three main parts are presented: a first one dedicated to the influence of the relaxation function setup, then consideration on achieved configurations for several MIP GAP values, the last one is focused on a sensitivity analysis considering various scenario: $CO_2$ tax, gas and electricity $CO_2$ content, presence or not of a wood boiler.

## 2 METHODOLOGY

### 2.1 Study case description

The system features a two-stage heating process (Figure 1a): a pre-heating stage, with a heat pump and a solar loop, and a main stage with two boilers. The cold branch flow, arriving at temperature $T_{cb}$, is initially pre-heated by the heat pump to a temperature $T_p$ before being further heated by the boilers to the set-point temperature $T_{hb}$. Bypass 2 line allows complete skipping of the boilers while Bypass 1 line allows either complete or partial skipping of the flow headed for the preheating stage. During Thermal Energy Storage (TES) charging, excess heat from the outlet of valve V6 is stored, while during discharging, the required heat is extracted and mixed with the main flow before distribution.

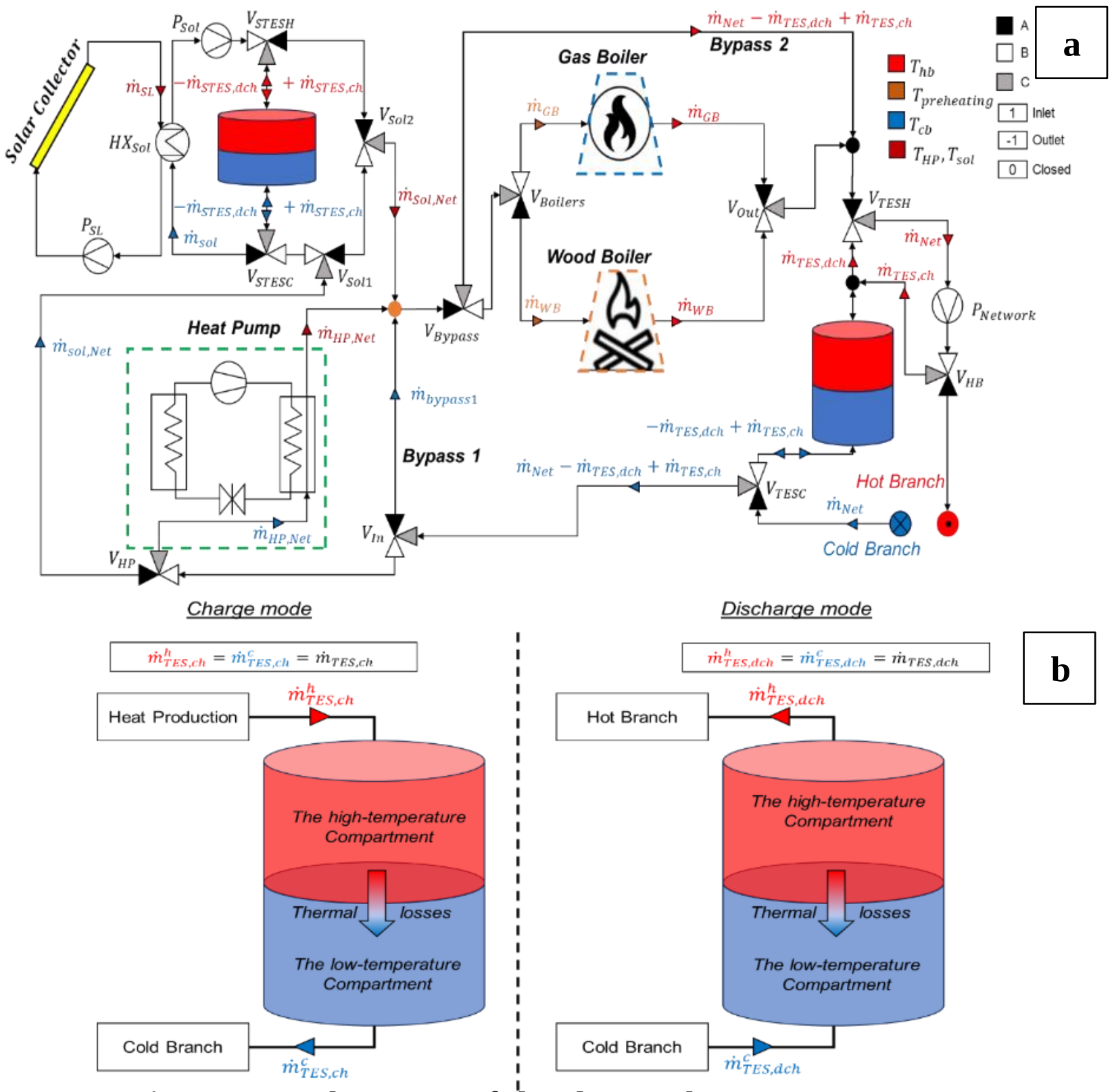

**Figure 1:** Schematic of the district heating system (a) and the thermal energy storage model (b)

For the solar loop, the setpoint of $\dot{m}_{sol}$ depends on the solar irradiation $GI$: 0 for $GI \leq 300\ W/m^2$, $\dot{m}_{sol,low} \geq 0.002\ kg/s.m^2$ for $300\ W/m^2 \leq GI < 700\ W/m^2$ and a $\dot{m}_{sol,high} \geq \dot{m}_{sol,low} + 0.002\ kg/s.m^2$ for $700\ W/m^2 \leq GI$. The wood boiler is set to be shut down during the summer period (from start of May to end of September).

The TES is modeled as a stratified tank with two compartments: a high-temperature ($T_{hb}$) and a low-temperature ($T_{cb}$) section (Figure 1b). It is designed to partially or fully cover the load for short periods. To simplify the model, the temperatures in both compartments are assumed to remain constant between charge and discharge cycles (Thomas et al., 2023). Thermal losses within the TES (and STES) are not modeled as external heat dissipation but rather as the gradual transformation of mass from the high-temperature compartment to the low-temperature compartment.

For heat load and outdoor temperature, the 2022 hourly data from Vaulx-en-Velin, France, are used. To avoid 2022 bias on energy prices due to the Russian-Ukrainian conflict, 2021 electricity prices and $CO_2$ emissions from the RTE platform were used ("Eco2mix," 2020). To optimize computational efficiency while maintaining accuracy, the dataset was condensed into 36 representative days grouped into 12 periods with 3-hour time steps.

## 2.2 Formulation of the optimization problem

The optimization problem is formulated as a Mixed Integer Linear Programming (MILP) with an economic and environmental criterion (LCOH, Levelized Cost of Heat) for the objective function as in equation (1) and (2):

$$\min LCOH = \frac{\sum_{\mathrm{n}=1}^{LCP} \frac{TAC}{(1+\mathrm{dr})^n}}{\sum_{\mathrm{n}=1}^{LCP} \frac{E_k}{(1+\mathrm{dr})^n}} \tag{1}$$

$$\mathrm{TAC} = CRF \cdot \left( \sum_{k=1}^{LD} \frac{C_{tot}^{inv}}{(1+dr)^k} + \sum_{k=LD+1}^{LCP} \frac{C_{tot}^{rep}}{(1+dr)^k} + \sum_{k=1}^{LCP} \frac{C_{tot}^{opt}}{(1+\mathrm{dr})^{\mathrm{k}}} + \sum_{k=1}^{LCP} \frac{C_{tot}^{CO_2}}{(1+\mathrm{dr})^{\mathrm{k}}} \right) \tag{2}$$

with $C_{tot}^{inv}$ is the total annual investment cost in euros, $C_{tot}^{opt}$ is the total annual operational cost in euros, $C_{tot}^{CO_2}$ is the total annual cost of $CO_2$ emissions in euros, $C_{tot}^{rep}$ is the total annual cost of replacement of installed equipment and CRF is the capital recovery factor. LD is the loan duration and LCP is the Life Cycle of the Project (40 years in this case).

The four costs are described in detail in the equations (2) to (5):

$$C_{tot}^{inv} = \sum_{j \in \{\mathrm{GB,WB,HP,Sol,TES,STES}\}} LF \cdot c_j^{inv} \cdot z_j \cdot P_j^{max}\ (or\ V_{TES}^{max}, V_{STES}^{max}) \tag{3}$$

$$C_{tot}^{opt} = \sum_{j \in \{\mathrm{GB,WB,HP,Sol}\}} \sum_{t \in \mathcal{T}} c_{j,t}^{opt} \cdot \Delta t \cdot P_{j,t}^{cons} \tag{4}$$

$$C_{tot}^{CO_2} = \sum_{j \in \{\mathrm{GB,WB,HP,Sol}\}} \sum_{t \in \mathcal{T}} c_{j,t}^{CO_2} \cdot \Delta t \cdot P_{j,t}^{cons} \tag{5}$$

$$C_{tot}^{rep} = \sum_{j \in \{\mathrm{GB,WB,HP,Sol,TES,STES}\}} LF \cdot \frac{c_j^{rep}}{(1+dr)^{L_j}} \cdot z_j \cdot P_j^{max}\ (or\ V_{TES}^{max}) \tag{6}$$

With:

$$CRF = \frac{dr\ \cdot\ (1+dr)^{LCP}}{(1+dr)^{LCP}-1} \tag{7}$$

$$LF = \frac{ir\ \cdot\ (1+ir)^{LD}}{(1+ir)^{LD}-1} \tag{8}$$

Where LF is the annuity factor and CRF the Capital Recovery Factor and $L_j$ the lifetime of equipment j.

Boilers: The performance curve of the gas boiler is taken as the average of the multiple curves from (Bannister, 2016; Short et al., 2017) and that of the wood boiler from (Świerzewski and Kalina, 2020).

To avoid non-linear constraint due to the non-linear performance curve, we use the method in (Voll, 2014) in order to form a linear approximation presented in equations (9) and (10).

$$P_{j,t}^{cons}\left(P_{j,t}^{prod},\ P_j^{max}\right) = a_j \cdot \ P_{j,t}^{prod} + b_j \cdot y_{j,t}.\ \ z_j \cdot P_j^{max}, \qquad j \in \{GB, WB\} \tag{9}$$

$$P_{j,t}^{prod} = \dot{m}_{j,t} \cdot C_p \cdot \left(T_{hb}\ - \left(T_{cb} + \Delta T_{ph,t}\right)\right), \qquad j \in \{GB, WB\} \tag{10}$$

In addition, the wood boiler must operate for at least three consecutive days after startup to maintain efficiency and reduce emissions and remain 12 hours off in case of a shutdown (Veyron et al., 2022) with limits on maximum part load modulation equal to 4.16% of maximum power :

$$\sum_{k=1}^{\mathrm{N_{WB,off}}-1} y_{WB,t+k} \leq \left(1 + y_{WB,t} - y_{WB,t-1}\right), \quad \forall t \tag{11}$$

$$\sum_{k=1}^{\mathrm{N}_{WB,,\mathrm{on}}-1} y_{WB,t+k} \geq y_{WB,t} - y_{WB,t-1}\ , \quad \forall t \tag{12}$$

$N_{off}$ is the required downtime and $N_{on}$ is the required uptime

<u>Heat-Pump & Solar thermal system:</u> The performance equation of the heat pump is derived from experimental curves (Ruhnau et al., 2019) and linearized to facilitate computational analysis as in equation (11).As for the performance equation of the solar thermal collectors, it is based on the model from (Gao et al., 2024) in equation (14). For linearity purpose, we neglect the second loss's factor

$$P_{HP,t}^{prod} = a_{HP} \cdot P_{HP,t}^{cons} + b_{HP} \cdot \Delta T_{HP,t} \cdot P_{HP,t}^{cons} + f_{HP}^{lin}(\Delta T_{HP,t}) \cdot P_{HP,t}^{cons} \tag{13}$$

$$P_{Sol,t}^{prod} = \eta_0 \cdot GI_t \cdot y_{Sol,t} \cdot A_{Sol} - a_{Sol} \cdot y_{Sol,t} \cdot A_{Sol} . \Delta T_{Sol,t} \tag{14}$$

$$\Delta T_{j,t} = T_{j,t} - T_{ext,t}, \qquad j \in \{HP, Sol\} \tag{15}$$

$$P_{j,t}^{prod} = \dot{m}_{j,t} \times C_p \times (T_{j,t} - T_{cb}), \qquad j \in \{HP, Sol\} \tag{16}$$

The preheating temperature is defined as the temperature of the mix of three flows: $\dot{m}_{sol}, \dot{m}_{HP}$ *and* $\dot{m}_{bypass1}$

$$\left(\dot{m}_{sol,t} + \dot{m}_{HP,t} + \dot{m}_{bypass1,t}\right) \cdot \Delta T_{ph,t} = \dot{m}_{sol,t} \cdot \Delta T_{Sol,t} + \dot{m}_{HP,t} \cdot \Delta T_{HP,t} \tag{17}$$

<u>TES (and STES):</u> Thus, the chosen method aligns with established practices in TES (and STES) modeling and is effective for optimizing energy system performance in applications requiring computational efficiency. From (Wirtz et al., 2021a), we consider the following TES model using the parameter mentioned by the authors.

$$m_{j,g(i),t+1}^{intra} = (m_{j,g(i),t}^{intra}) \times (1 - \phi) + \Delta t . \eta_{ch} . \dot{m}_{j,ch,t} - \frac{\Delta t}{\eta_{dch}} . \dot{m}_{j,dch,t}, \quad j \in \{TES, STES\} \tag{18}$$

$$m_{j,i+1}^{inter} = (m_{j,i}^{inter} - m_{j,g(i),0}^{intra}) \times (1 - \phi)^\wedge N + m_{j,g(i),N}^{intra}, \quad j \in \{TES, STES\} \tag{19}$$

$$P_{j,t}^{ch\ (or\ dch)} = \dot{m}_{j,ch\ (or\ dch),t} \times C_p \times \Delta T_{max}\ , \quad j \in \{TES, STES\} \tag{20}$$

$m_{g(i),t}^{intra}$ presents the mass of high temperature compartment's mass in TES (and STES) at the instant $t$ during representative period g(i), $m_i^{inter}$ presents the mass of hot water in TES (and STES) at the beginning of period $i$ and $\dot{m}_{TES,ch,t}$ *and* $\dot{m}_{TES,dch,t}$ present the charge and discharge mass flow respectively. $N$ is the number of timesteps within the representative period g(i), $\phi$ is the loss factor, $\Delta t$ is the duration of charge/discharge and $\eta_{ch}$ and $\eta_{dch}$ are respectively the charge and discharge efficiencies. For the case of the STES, the charge is limited to the summer period and discharge is limited to the rest of the year. Table 1 summarizes the energy and mass balance constraints, along with the requirement of 50% of the total production must be sourced from renewable-based heat sources, including solar-based heat, wood-based heat, and non-electric heat pump-based heat.

### 2.3 Linearization of the optimization problem

The first non-linear constraint involves the multiplication of binary and continuous variables. To manage this, we apply the linearization technique outlined in (Asghari et al., 2022) which introduces a new continuous variable. The second non-linear constraint addresses temperature variations at the preheating stage. To represent these temperatures in a linearized model, we employ a discretization approach in (Asghari et al., 2022): A set $S = \{ 2^i, 0 \le i \le NT \}$ with $NT = \min_n \{2^{n+1} \ge \Delta T_{max} = T_{hb} - T_{cb}\}$. A binary combination of values in $S$ approximate a given temperature with discretization step equal or larger than 1: $\Delta T = \sum_0^{NT} \alpha_i . 2^{\mathrm{i}}$ where $\alpha_i$ are binary variables.

**Table 1**: Energy and mass balance constraints

| Description | Constraint | N° |
|---|---|---|
| Energy balance constraint | $\sum_{c\in\{GB,WB,HP,Sol\}} \mathrm{P}_{c,t}^{prod} + \mathrm{P}_{TES,t}^{Dech} - \mathrm{P}_{TES,t}^{Char} + \mathrm{P}_{STES,t}^{Dech} - \mathrm{P}_{STES,t}^{Char} = \mathrm{HL}_t, \quad \forall t$ | *(21)* |
| Delivery point-mass balance constraint | $\dot{m}_{GB,t} + \dot{m}_{WB,t} = \left( \frac{\mathrm{HL}_t}{\mathrm{C_p} \times \Delta\mathrm{T}} + \dot{m}_{TES,\mathrm{dch,t}} - \dot{m}_{TES,ch,\mathrm{t}} + \dot{m}_{STES,\mathrm{dch,t}} \right), \forall t$ | *(22)* |
| Preheat point-mass balance constraint | $\dot{m}_{HP,t} + (\dot{m}_{Sol,t} - \dot{m}_{STES,ch,t}) = \left( \frac{\mathrm{HL}_t}{\mathrm{C_p} \times \Delta\mathrm{T}} + \dot{m}_{TES,\mathrm{dch,t}} - \dot{m}_{TES,ch,\mathrm{t}} + \dot{m}_{STES,\mathrm{dch,t}} \right), \forall t$ | *(23)* |
| 50% Renewable production rate constraint | $\sum_{t=1}^{T} \mathrm{P}_{WB,t}^{prod} + (\mathrm{P}_{HP,t}^{prod} - \mathrm{P}_{elec,t}) = 50\% \cdot \sum_{t=1}^{T} \sum_{c\in\{GB,WB,HP\}} \mathrm{P}_{c,t}^{prod} + \mathrm{P}_{TES,t}^{Dech} - \mathrm{P}_{TES,t}^{Char}$ | *(24)* |

The economic parameters are given in Table 2, where data were taken from ("Technology catalogues," 2016) (Sifnaios et al., 2023)

**Table 2**: Economic and technical parameters for technologies

| | Investment cost | Fixed O&M costs | Variable O&M costs | Lifetime |
|---|---|---|---|---|
| Gas boiler | 60 k€/MW | 2073.58 €.MW$^{-1}$.year$^{-1}$ | 1.17 €/MWh | 20 years |
| Wood boiler | 520 k€/MW | 44661 €.MW$^{-1}$.year$^{-1}$ | 2.87 €/MWh | 20 years |
| Solar thermal | 0.198 k€/m$^2$ | 40 €.m2$^{-1}$.year$^{-1}$ | 0.22 €/MWh | 20 years |
| Heat pump | 1,010 k€/MW | 2212.6 €.MW$^{-1}$.year$^{-1}$ | 2.33 €/MWh | 20 years |
| TES | 0.250 k€/m$^3$ | 0.47 €.m3$^{-1}$.year$^{-1}$ (2% of Inv) | 0 | 40 years |
| STES | 0.120 k€/m$^3$ | 0.282 €.m3$^{-1}$.year$^{-1}$ (2% of Inv) | 0 | 40 years |

# 3 RESULTS AND DISCUSSIONS

### 3.1 The simulations' scenarios and MIP convergence

To analyze the impact of different carbon emission scenarios on the design of the DHS, we defined the 6 following cases in which a combination of them defines different scenarios:

- Wood boiler : No Wood Boiler "NWB" and With Wood Boiler "WWB".
- Carbon tax : "CT74" (Carbon Tax = 74 euros/$t_{CO2}$) and "CT200" (200 euros/$t_{CO2}$).
- Electricity $CO_2$ content : "RE" (2021 French mix carbon footprint with an average of 37.94 $kg_{CO2}$/MWh) and "HE" (2021 Greman mix carbon footprint, with HE ~ 8× RE).

- Gas $CO_2$ content: "NG" (Natural Gas ~ 227 $CO_2$eq/kWh) and "BG" (Biogas ~ 22 $CO_2$eq/kWh).

The considered reference scenario is "NWB-CT74-RE-NG".

For scenarios with "NWB", the model struggles with operating the solar thermal system due to limitations on the set-point temperature. The difficulty is to cope with fluctuating solar irradiations and a limited set-point temperature while operating on limited design mass flowrates ($\dot{m}_{sol,low}$ and $\dot{m}_{sol,high}$) which results in more than a day of calculation to reach a 19% MIP GAP with high reliance on natural gas during summer in order to avoid a low design mass flowrate that can cause overheating during peak of irradiations (the as seen if figure 2b, the flowrates are high which results in preheating temperature (86.5°C) lower than the setpoint temperature and the rest is fulfilled with Gas boiler). To counter this issue, a relaxation variable is added to the model allowing the solar thermal to reach high temperatures but discarding the surplus into the environment to avoid an overheating.

$$P_{Sol,t}^{prod} = \eta_0 \cdot GI_t \cdot y_{Sol,t} \cdot A_{Sol} - a_{Sol} \cdot y_{Sol,t} \cdot A_{Sol} \cdot \Delta T_{Sol,t} - P_{relax,t} \tag{25}$$

$$0 \leq P_{relax,t} \leq \dot{m}_{sol,t} \cdot C_p \cdot \Delta T_{sol,relax} \tag{26}$$

With $\Delta T_{sol,relax}$ is the maximum surplus temperature.

As seen in Figure 2a, the relaxation accelerates the convergence of the problem from a 19% MIP GAP in 26 hours to 10% MIP GAP in 12 hours with a dissipation of 6% of the total solar production. The higher the maximum surplus, the more the model dissipates the energy but ensure a more fast and stable convergence. The figure 2b shows the production of each generator during the summer. The relaxation reduces the solar $\dot{m}_{sol,low}$ (SMF low) by 24% and the solar $\dot{m}_{sol,high}$ (SMF high) by 8% allowing an overheating over the set-point temperature which results on relying less on the gas boiler by 77% and the heat pump by 60%.

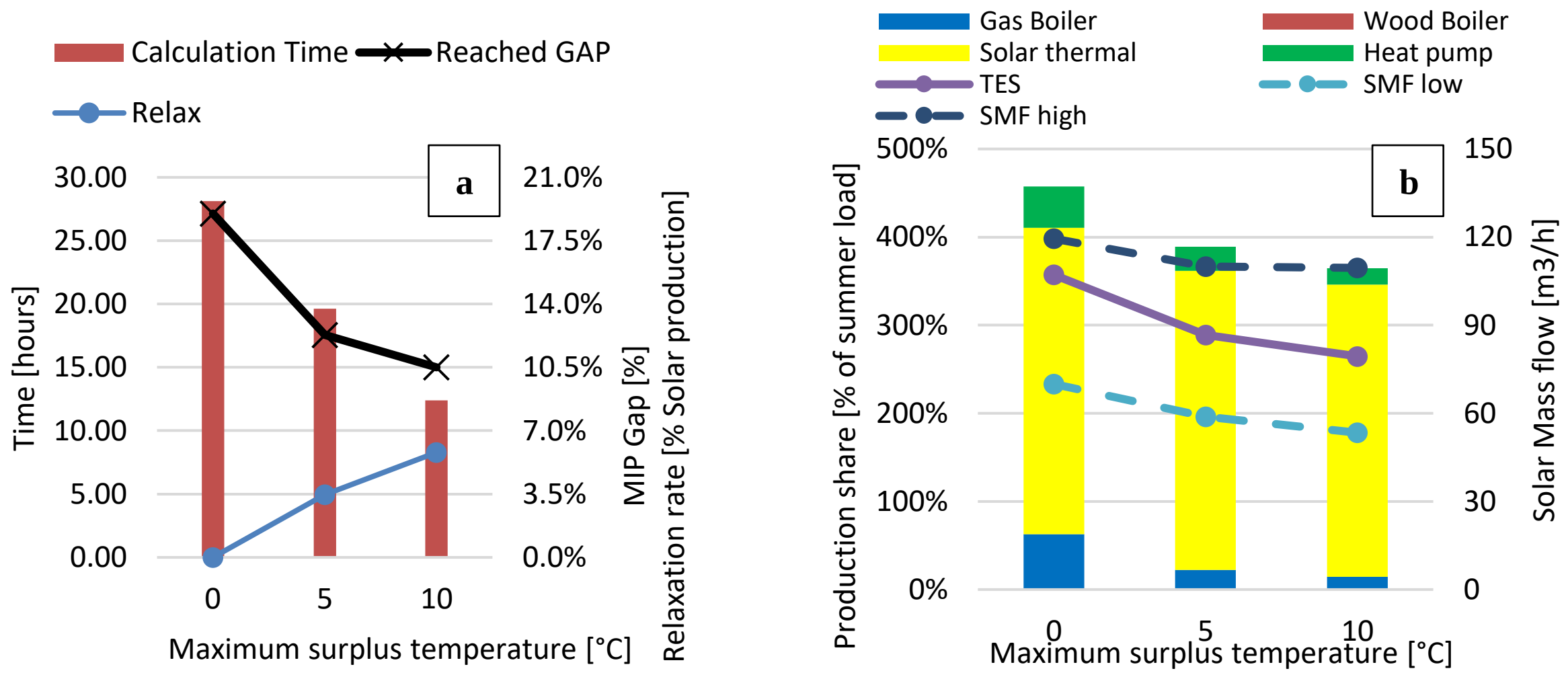

**Figure 2:** Reached MIP GAP (a) and production share during summer (b) in function of the maximum surplus temperature

The MIP GAP evolution of other "NWB-CT74" scenarios with different gas and electricity $CO_2$ emissions is plotted in Figure 3a. A MIP GAP of 10% is deemed a safe choice giving the convergence starts to slow down after 15% (on average it takes 8 hours for 4% decrease in gap). To further analyse the different MIP GAP impact on the DHS design and operation, Figure 3b and Figure 3c show the evolution of the design and operations of the DHS at each MIP GAP.

At GAP = 30%, the system is dominated by the heat pump with an installed capacity of 7.3 MW, while the gas boiler remains low at 1.5 MW. No solar thermal is utilized, and TES capacity is only 1600 $m^3$, indicating minimal reliance on storage. The heat pump provides nearly the totality of the demand along with minor losses from TES (12%) while the gas boiler provides nearly 2%. LCOH is high at 83.96 euros /MWh, reflecting the high operational cost of a system relying heavily on electricity-driven heating.

At GAP = 25% (time = 4.39 hours), the system integrates solar thermal (3100 m2) and expands TES to 2660 m3, increasing storage flexibility. The heat pump capacity remains high at 7.2 MW, but the gas boiler increases significantly to 3.1 MW, covering 14.63% of the load. Solar thermal contributes 12.3%, leading to a shift towards mixed energy sources. TES losses increase to 23%, suggesting higher reliance on storage cycling. The LCOH slightly drops to 82.2 euros /MWh, showing early cost improvements due to diversification.

As the GAP reduces to 13% (time = 6.07 hours), the system continues shifting towards solar integration. Solar thermal expands to 8100 m2, while the heat pump reduces to 4.6 MW. The gas boiler reaches 4 MW, covering 33.11% of the demand, indicating a further shift toward fossil-based heating. TES increases to 6200 $m^3$, supporting solar energy integration, but TES efficiency declines, as shown by 49 % losses. This configuration significantly reduces LCOH to 70.44 euros/MWh, demonstrating the economic impact of storage expansion and solar integration.

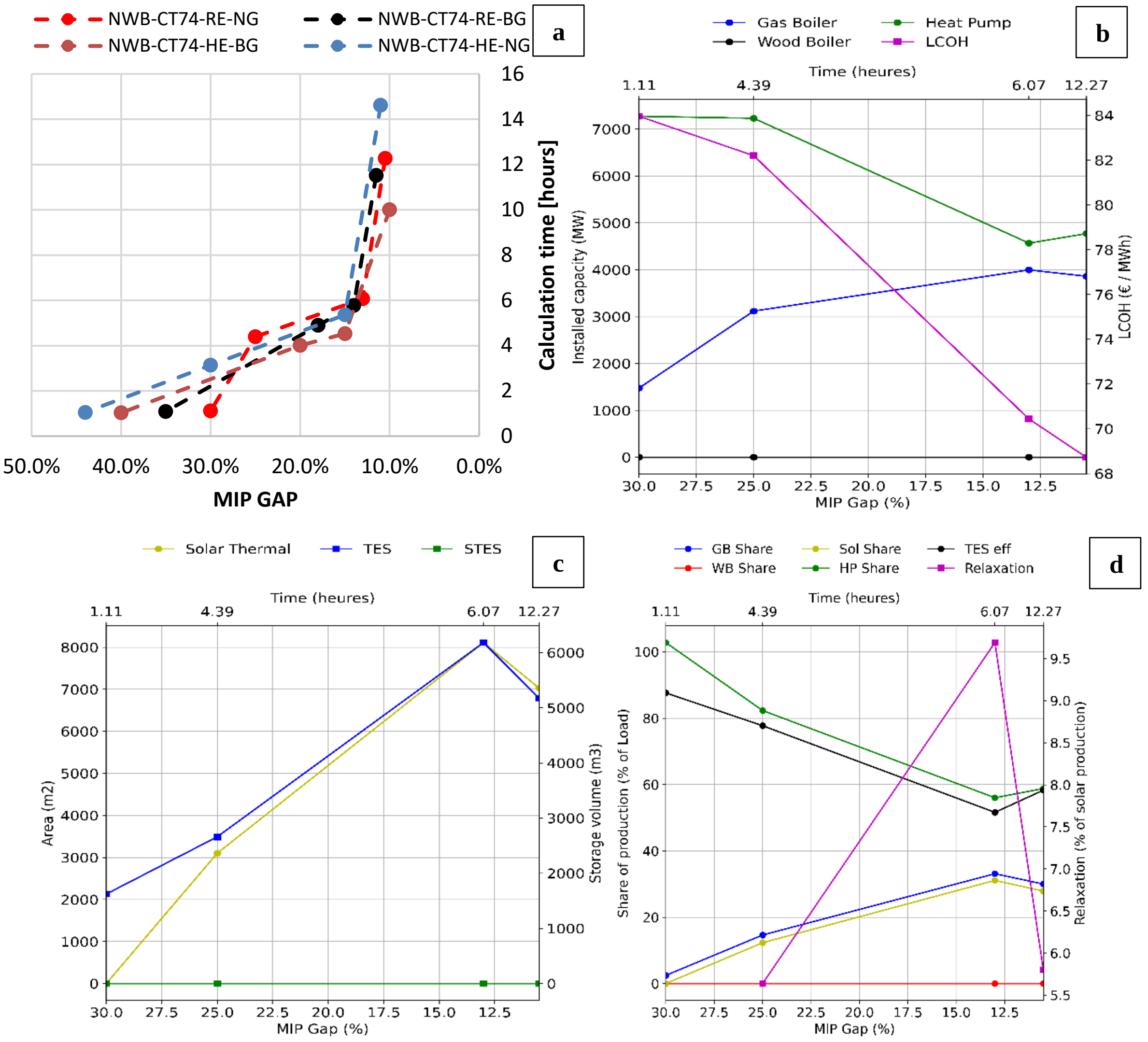

**Figure 3:** Reached MIP GAP for NWB-CT74 scenarios in function of calculation time (a) the installed capacity of the GB, WB, HP and LCOH (b) The installed capacity of the solar thermal, TES and STES (c) The share production of the heat generators and the storage efficiency (d) in function calculation time and MIP GAP for reference scenarios

At GAP = 10.5% (time = 12.27 hours), the system reaches an optimized balance. Solar thermal remains stable at 7030 $m^2$, while TES slightly reduces to 5170 $m^3$, optimizing storage efficiency. The heat pump reduces to 4.7 MW, while the gas boiler stabilizes at 3.9 MW, supplying 30% of the demand. The LCOH further decreases to 68.73 euros/MWh, marking the lowest cost achieved. TES losses stabilize at 42 %, indicating an efficient level of storage utilization. Overall, as the GAP decreases from 30 % to 10.5 %, the system transitions from a heat pump-dominant setup with high costs and minimal storage, to a balanced configuration with integrated solar thermal, TES, and gas boilers, leading to significant LCOH reductions. The economic benefits of a diversified system become evident at GAP = 13%, with diminishing returns beyond GAP = 10.5%. Beyond this point, further refining the GAP mainly redistributes capacities without introducing major structural changes, leading to diminishing cost improvements.

### 3.2 Reference case and multi-scenario analysis: DHS's design and operation

The reference case does not include a wood boiler but still have to ensure 50% of the heat production is renewable-source based. The solution reached a configuration of 3.9 MW of gas boiler, 4.7 MW for the heat pump, 7030 m2 for the solar thermal and 5170 m3 of TES. The evolution vs. time of the heat production by generator and the that of the demand are plotted in Figure 4, with appearing of the hot/cold seasons: first cold season (from start of January to end April), summer season (from start of May to end of September) and second cold period (from start of October to end of December).

The first cold period is characterized by low electricity prices (average of 60 euros/MWh) and medium $CO_2$ emissions (average of 44 $t_{CO2}$/MWh). The heat pump operates at a temperature around 80°C consuming electricity with a COP fluctuating between 2 and 2.2. It is complemented with the gas boiler due to the peak of demand with "pushes" for the solar system to reach the set-point temperature. During the second period, the demand decreases with total shutdown of the gas boiler, preheating technologies already achieving the setpoint for the operating temperature. The heat pump remains main producer at temperature equal to the set-point temperature of 90°C while solar thermal mostly fill the TES to counter hourly fluctuation. During the third period, the heat pump starts to operates based on low-electricity prices (for an electricity price around 14 euros/MWh and CO2 emission of 20 kg$CO_2$/Mwh in the first day before they rise in the third day to 60 euros/MWh and 47 kg$CO_2$/MWh) while relying on storage management to fulfill the load.

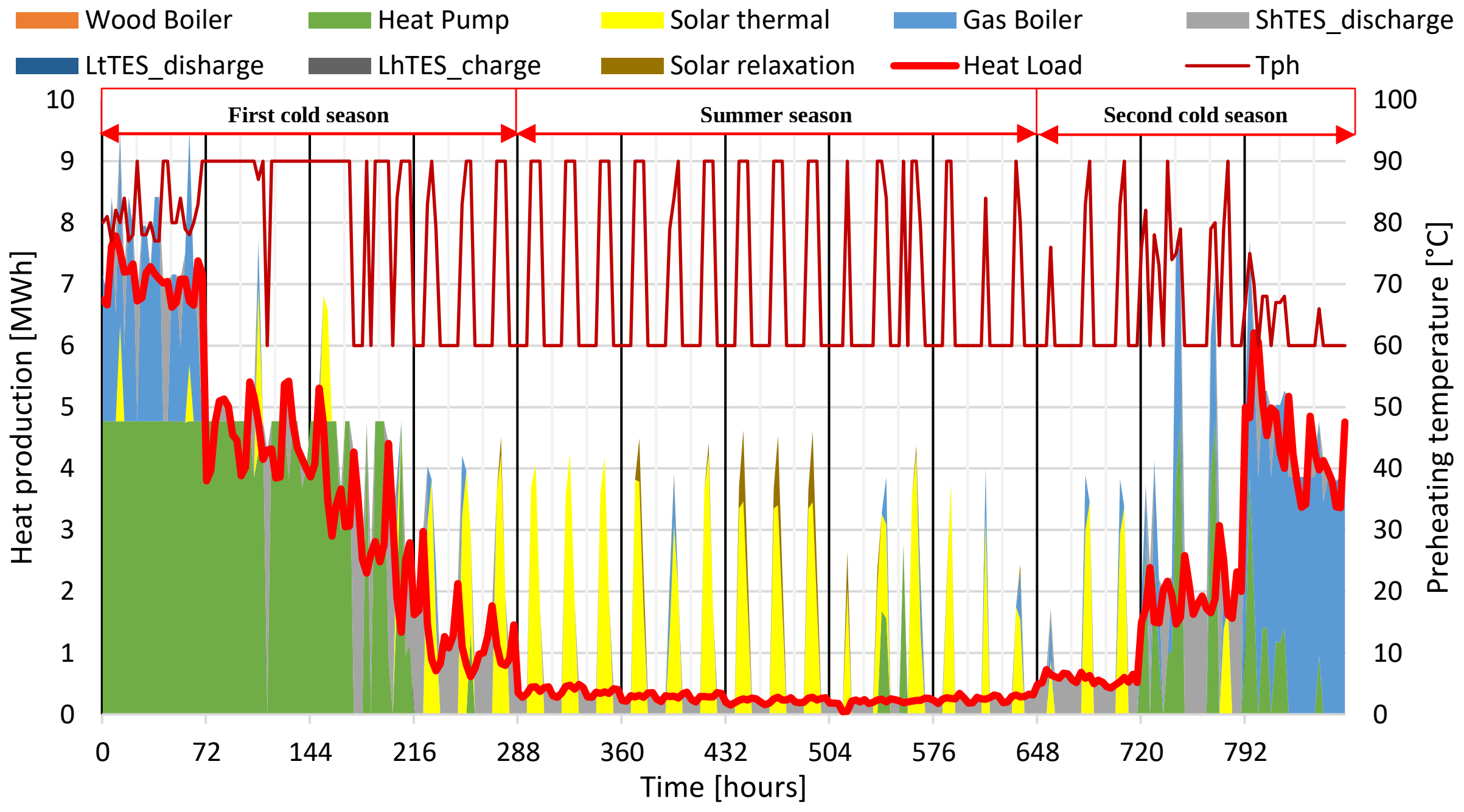

**Figure 4:** Operation schedule and preheating temperature profile for the reference case

The summer season is characterized by high solar irradiations ($\geq 600\ W/m^2$). The DHS operates totally on solar thermal energy while relying on the TES to provides heat during the nights. The relaxation during this period, specifically in July, plays an important role to avoid overheating the solar thermal with 6% of solar production is lost to the environment along with the rest of thermal losses. The gas boiler and heat pump still chip in few hours during low irradiations (between 400 $W/m^2$ and 550 $W/m^2$) at the end of summer period. The second cold season starts with emptying the full TES along with solar production. This season is characterized by high electricity prices and CO2 emissions (252 euros/MWH and 61 $t_{CO2}$/MWh respectively). The heat pump contributes at the beginning along with gas boiler while taking advantage of the Solar energy stored in the TES. Due to the jump, in electricity prices and the solar irradiations below the required minimum ( 300 $W/m^2$), The gas boiler become the only profitable and available solution during December with preheating around 65°C using the heat pump during the drops in prices.

To analyze the impact of the different scenarios on the design and operation of the DHS, multiple simulations are run using the same procedure explained for the refence case. In the figures 5 and 6, the reference case; RE-NG, is always the first histogram of the figure. Figure 5a shows that in NWB-CT74, gas boilers installed capacity is 3.86 MW, 4.76 MW for heat pumps, and 7,029 m² solar field. Increasing the carbon tax and switching to biogas (NWB-CT200, figure 5b) enhances solar integration to 13,370 m², while heat pump capacity drops to 2.74 MW due to increased reliance on solar thermal and TES. Without the wood boiler, the system must compensate for seasonal variations, leading to a higher TES requirement and greater operational flexibility.

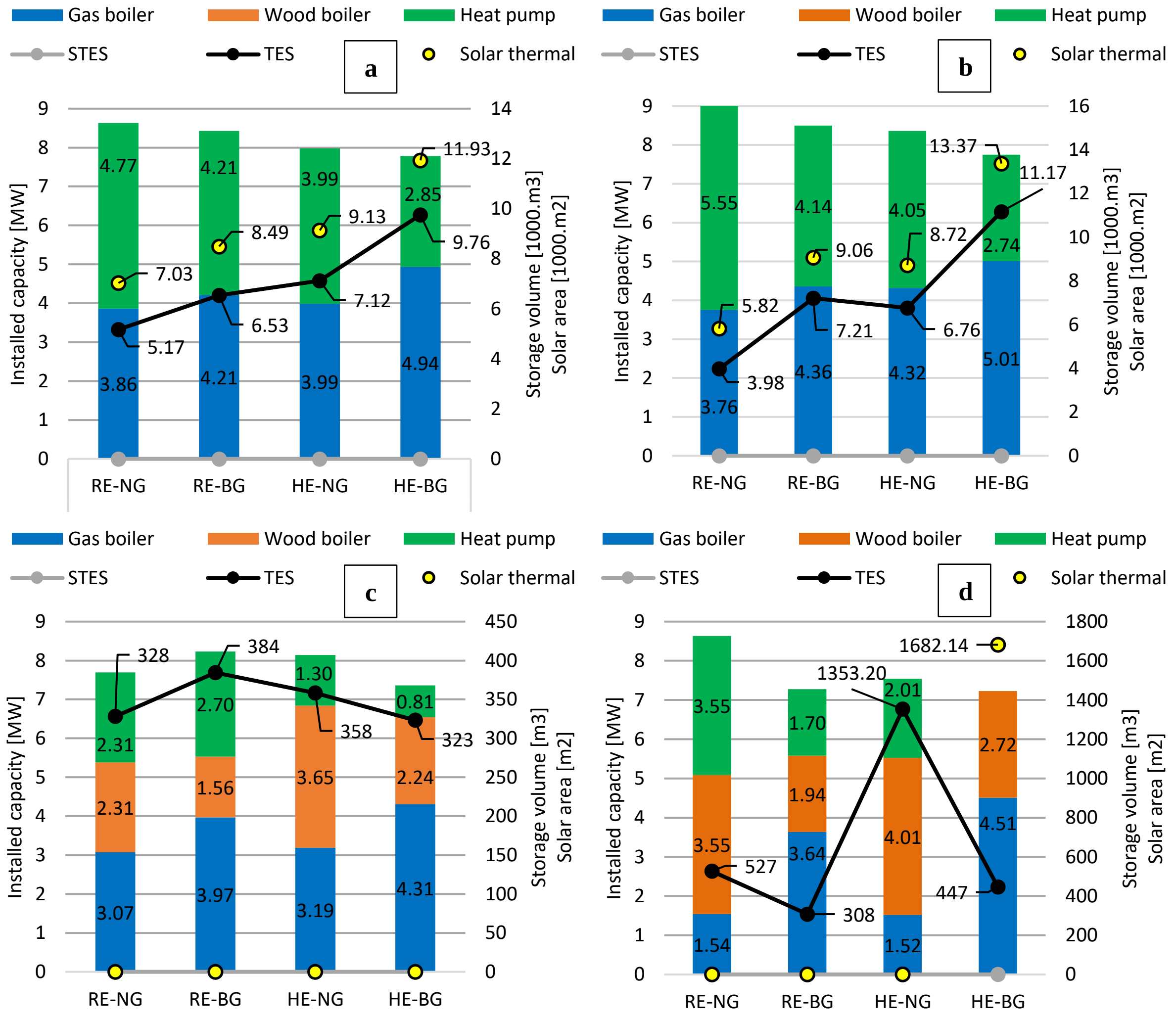

**Figure 5:** The different design's configurations for scenarios (a) NWB-CT74, (b) NWB-CT200, (c) WWB-CT74, (d) WWB-CT200.

When the wood boiler is included (WWB), the system shifts towards combustion-based heating, reducing dependence on TES and solar thermal. WWB-CT74 (figure 5c) installs a 2.31 MW wood boiler, eliminating the need for solar thermal, while WWB-CT200 (figure 5d) increases wood boiler capacity to 3.55 MW, reducing heat pump and TES requirements due to carbon emissions. The increased carbon tax impacts NWB by promoting solar deployment but has a limited effect on WWB, where biomass remains dominant.

Figure 6 illustrates operational shares and TES efficiency. In NWB-CT74, gas boiler contributes 30%, heat pump 30%, and solar thermal 27.8%. In NWB-CT20-HEBG, solar thermal rises to 53% at the expense of heat pumps, while gas boiler reliance increases to 55% , due to biogas consumption, to balance intermittency. TES cycling becomes more prominent, with losses rising from 42% to 65% as the model compensates for solar fluctuations. In WWB-CT74, wood boilers provide 45% of heat, reducing reliance on TES and solar thermal. Under WWB-CT200, wood boilers dominate production, leading to minimal TES cycling and lower efficiency losses. TES plays a critical role in NWB by stabilizing solar-driven variations, whereas WWB configurations operate with continuous combustion, reducing the need for storage. Figure 7 highlights preheating temperature (Tph) variations. In NWB-CT74, Tph averages 80°C in winter, increasing to 84°C in NWB-CT200 due to greater solar integration and TES cycling. The solar thermal system supplies temperatures around 60°C under normal conditions but peaks at 84°C in NWB-CT200, necessitating dissipation measures to prevent overheating. Heat pump operates between 75°C and 84°C, adapting to electricity price fluctuations.

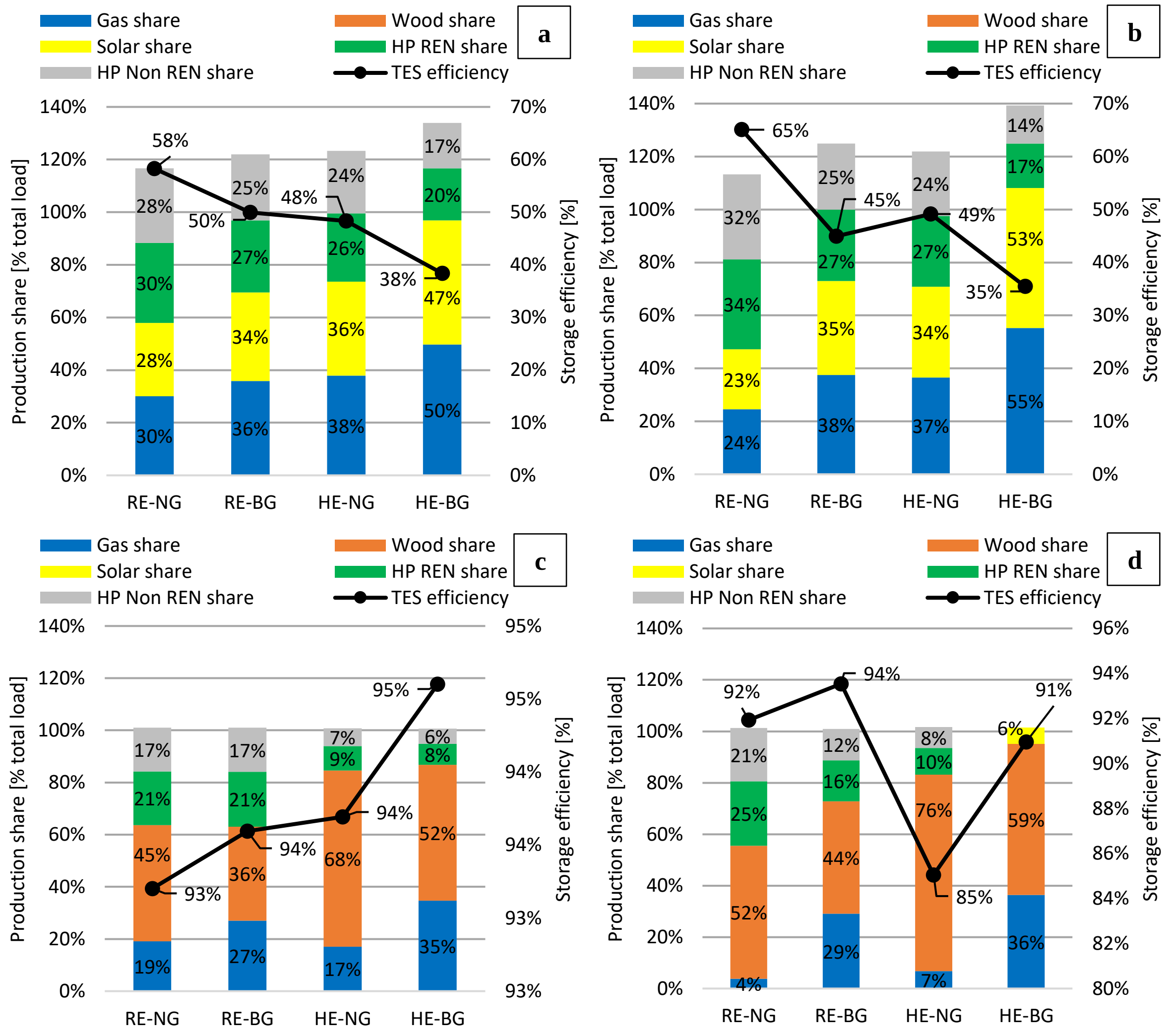

**Figure 6:** The different production shares and TES efficiencies for scenarios (a) NWB-CT74, (b) NWB-CT200, (c) WWB-CT74, (d) WWB-CT200.

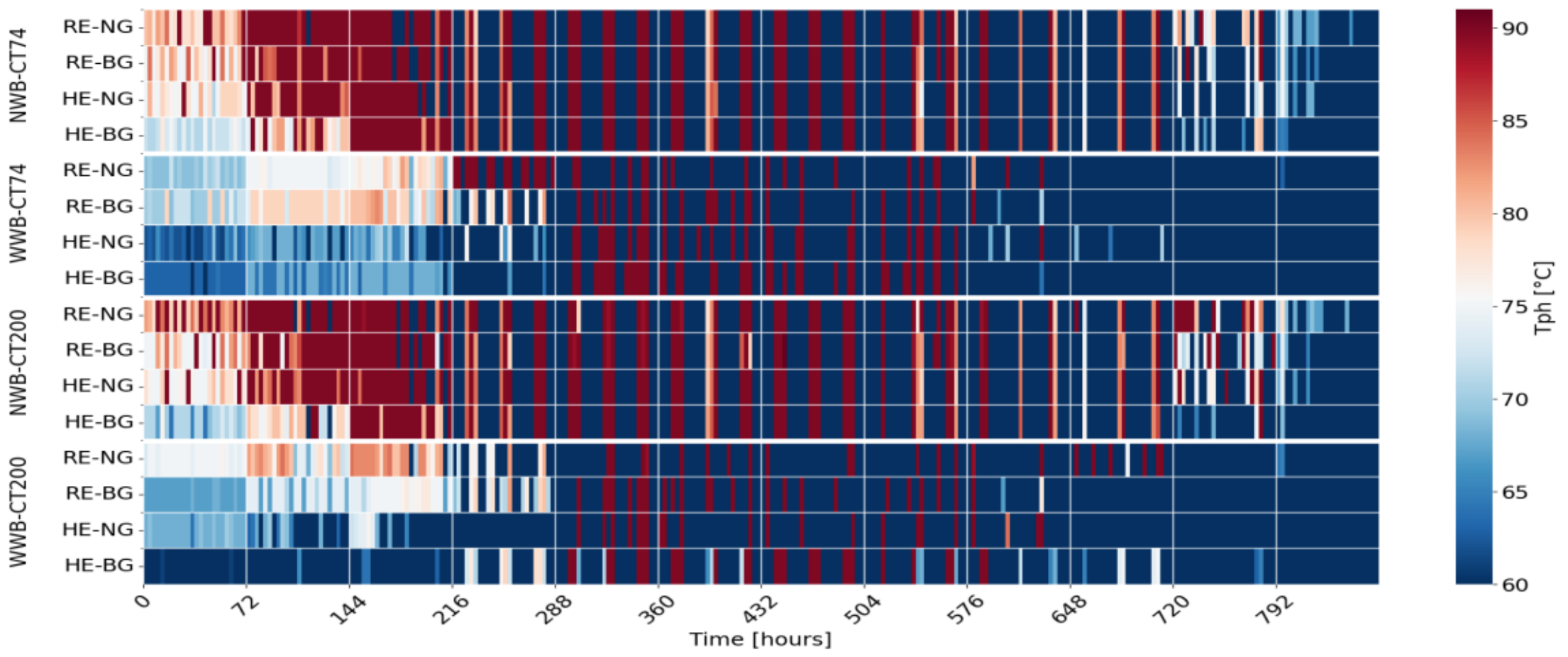

**Figure 7:** Preheating temperature profiles for the different scenarios.

During summer, solar thermal dominates, but its efficiency depends on TES utilization to avoid temperature surges. In WWB-CT74, Tph stabilizes at 69–72°C due to continuous wood combustion, ensuring steady heat production without requiring TES cycling. WWB-CT200 further stabilizes Tph by reducing TES dependency and relying more on direct combustion, maintaining temperatures consistently below 72°C. NWB scenarios experience wider temperature fluctuations due to solar and TES reliance, whereas WWB maintains a more consistent temperature profile through biomass combustion.

## 4 CONCLUSIONS

In the present paper, a new MILP approach including a relaxation function to optimize the design of DHS based on temperature dependent technologies (heat-pump and solar thermal panel field) is proposed. A scenario allowing a combination of natural gas boiler, heat-pump and solar thermal panels + storage is used as reference. The influence of the carbon tax, the electricity/gas carbon content and presence or not of a wood boiler has been studied. The main results are resumed hereafter:

- The setup of the relaxation function heavily impacts the achieved configuration as well as the calculation time. The best trade-off was found for a setup at 10 °C which allows significantly the calculation time while allowing solar thermal loop (e.g. for low setup values, the setpoint temperature of the solar loop is nearly never reached because too strict constraints during summer overheating) to be installed without dropping too much excess solar heat,
- The MIP GAP at which the optimization is stopped, has a strong influence on achieved configuration, e.g. solar integration always requires a MIP GAP below 13%. The best trade-off was found to be around 10% to keep reasonable calculation time (< 15h),
- For scenarios without wood boiler, the $CO_2$ tax strongly influences the optimal configuration for the reference case (low carbon electricity + natural gas boiler) with example an increase by 25% of the HP installed capacity while the solar panels area decreases by 17%, due to reliance on gas boiler, when shifting from 74 to 200€/$t_{CO2}$,
- For scenarios with wood boiler, the solar energy is discarded for all cases excepted for electricity with high carbon content combined with high CO2 tax and low carbon content gas, the solar being still dependent on a presence of a gas boiler.

### NOMENCLATURE

**Parameters and variables**

| | | |
|---|---|---|
| dr | Discount rate | |
| ir | Loan interest rate | |
| $z$ | Binary design variable | |
| $y$ | Binary operation variable | |
| $P_j^{max}$ | the installed capacity of technology | (MW) |
| $P_j^{prod}$ | the installed capacity of technology | (MWh) |
| $P_j^{cons}$ | the installed capacity of technology | (MWh) |
| $A_{sol}^{max}$ | the area of installed Solar collector | ($m^2$) |
| $C_p$ | The HTF thermal capacity | (kJ/(kg.°K)) |
| $E_n$ | total yearly heat load | (MWh) |

**Subscript**

| | |
|---|---|
| TES | Thermal Energy Storage |
| STES | Seasonal Thermal Energy Storage |
| n | Year index |
| t | Timestep index |
| j | Technology index |

## ACKNOWLEDGEMENT

This work was supported as part of the CORRES project funded by Institut Carnot Ingénierie@Lyon.